\documentclass[aps,pra,reprint,showpacs,amssymb,amsmath,superscriptaddress]{revtex4-1}

\usepackage{lipsum}

\usepackage{graphicx}
\usepackage{color}
\usepackage{times}
\usepackage{ulem}
\usepackage{verbatim}
\usepackage{siunitx}
\usepackage{booktabs}

\newcommand{\mean}[1]{\expec{#1}}
\newcommand{\Rb}{$^{87}$Rb }

\newcommand{\CS}{\mathrm{CS}}
\newcommand{\sigsum}{\Sigma_S}
\newcommand{\angsum}{\Sigma_\theta}
\newcommand{\AU}{Institut for Fysik og Astronomi, Aarhus Universitet, Ny Munkegade 120, 8000 Aarhus C, Denmark.}

\newcommand{\expec}[1]{\langle #1 \rangle}

\newcommand{\comm}[1]{}
\newcommand{\Deff}{\Delta_\mathrm{eff}}

\newcommand{\Nnorm}{\mathcal{N}}
\newcommand{\Nnormone}{\mathcal{N}_1}
\newcommand{\Nnormtwo}{\mathcal{N}_2}

\DeclareMathOperator{\Var}{Var}

\begin{document}
\title{Sub-atom shot noise Faraday imaging of ultracold atom clouds}

\author{M. A. Kristensen}\email[Corresponding author: ]{mick@phys.au.dk}
\affiliation{\AU}
\author{M. Gajdacz}
\affiliation{\AU}
\author{P. L. Pedersen}
\affiliation{\AU}
\author{C. Klempt}
\affiliation{Institut f\"ur Quantenoptik, Leibniz Universit\"at Hannover, Welfengarten 1, D-30167 Hannover, Germany.}
\author{J. F. Sherson}
\affiliation{\AU}
\author{J. J. Arlt}
\affiliation{\AU}
\author{A. J. Hilliard}
\affiliation{\AU}

\begin{abstract}
We demonstrate that a dispersive imaging technique based on the Faraday effect can measure the atom number in a large, ultracold atom cloud with a precision below the atom shot noise level. The minimally destructive character of the technique allows us to take multiple images of the same cloud, which enables  sub-atom shot noise measurement precision of the atom number and allows for an \textit{in situ} determination of the measurement precision. We have developed a noise model that quantitatively describes the noise contributions due to photon shot noise in the detected light and the noise associated with  single atom loss. This model contains no free parameters and is calculated through an analysis of the fluctuations in the acquired images. For clouds containing $N\sim5\times 10^6$ atoms, we achieve a precision more than a factor of two below the atom shot noise level.
\end{abstract}

\maketitle
\section{Introduction}\label{sec:Introduction}

Dispersive imaging is the standard method to perform spatially resolved \textit{in situ} measurements on trapped ultracold atom clouds.
The atomic density distribution  gives rise to a spatially varying refractive index for  off-resonant imaging light and the dispersive imaging technique converts the acquired phase shift into an intensity modulation that can be measured on a scientific camera.
Imaging with off-resonant light   permits a sample to be probed several times due to the low rate of spontaneous emission. 
This minimally destructive character of the imaging is typically used to measure the cloud's evolution as it undergoes some dynamic process, such as the formation of a Bose Einstein condensate (BEC)~\cite{Andrews1996,Bradley1997,2016arXiv160104425W}. It has also been used to measure the speed of sound ~\cite{PhysRevLett.79.553} and to directly observe  vortices \cite{PhysRevLett.83.2498} in a Bose gas. Most recently, dispersive imaging was used to observe the quantum enhancement of the refractive index in a BEC~\cite{PhysRevLett.116.173602}.

Here, we investigate the use of repeated dispersive images to precisely determine the number of atoms in an ultracold thermal cloud.
We make use of up to 100 images of a single  cloud to enhance the measurement of the atom number. Another significant benefit of this approach comes from analysing the statistics of the repeated measurements, which provides an `in-measurement' characterisation of the imaging noise. There are two fundamental noise sources that limit the precision in determining  the atom number: the photon shot noise on the detected light and the noise arising from atom loss from spontaneous photon scattering induced by the imaging light. We develop a noise model that quantitatively describes these noise contributions. This  model contains no free parameters and is based on an analysis of the fluctuations in the acquired images.  For  clouds containing $N\sim5\times10^6$ atoms, a single dispersive image achieves a precision of approximately four times the atom shot noise level $\sqrt{N}$. When averaging over only 10 images of the same cloud, our imaging achieves sub-atom shot noise precision. 

The dispersive imaging technique we employ is based on the Faraday effect, whereby the linear polarisation plane of an off-resonant imaging beam  is rotated due to the vector part of the atomic polarisability. In earlier work, we compared the signal-to-noise ratio for four common dispersive imaging techniques \cite{Gajdacz2013}: phase contrast ~\cite{PhysRevLett.79.553} and dark field imaging~\cite{Andrews1996}, which typically make use of the scalar part of the atomic polarisability, and two variants of Faraday imaging~\cite{PhysRevA.55.3951, Kaminski2012}. 

However, our experiments and analysis investigate a fundamental feature of any dispersive imaging method: the interplay between the acquired phase shift  and the concomitant  spontaneous photon scattering that must occur \cite{Hope2004}. In this sense, dispersive imaging of a non-degenerate gas occupies an interesting position between the paradigms of  classical and quantum measurements: the trapped  cloud initially contains exactly $N_0$ atoms, but the imaging induces loss in the cloud through spontaneous photon scattering, causing a decrease in $N$ and an increase in the atom number uncertainty $\Delta N$ due to the stochastic character of the loss process.

The paper is structured as follows. In the following section, our realization of Faraday imaging is outlined, and in section \ref{sec:ExpImplementFaradatImg} we give details on the experimental setup. Section \ref{Image evaluation} describes the  processing we perform on the raw data to achieve high quality Faraday images. We also discuss the steps we take to obtain  accurate atom numbers through a cross-calibration of Faraday imaging with the more common absorption imaging. In section \ref{sect:Precision}, we analyse the achieved precision in the determination of the atom number in a single Faraday image. We then extend this analysis to the case of averaging over several images. We offer conclusions on our work in section \ref{sect:Conclusion}.

\section{Dark Field Faraday Imaging}\label{sec:DFFI}
Faraday imaging exploits circular birefringence in spin-polarised atomic samples: a linearly polarised off-resonant beam of light may be decomposed into an equal superposition of left and right hand circular polarisations; as the beam passes through the atomic sample,  the light-matter interaction leads to a differential phase shift of the two circular   components of the light leading to a rotation of the  linear polarisation plane. For an ultracold cloud of \Rb  in the ${|F=2, m_F = 2 \rangle}$ state that is spin polarised along the $z$ axis and probed with light close to the D2 transition,  the Faraday rotation angle is given by 

\begin{equation}\label{eq:theta}
\theta(x,y) =\frac{\hbar m_F\Gamma \lambda^2 }{16\pi \Deff}\int n(x,y,z)dz\equiv c_\textrm{F}(\Deff) \tilde{n}(x,y), 
\end{equation}
where $n(x,y,z)$ is the atomic density distribution,  
\mbox{$\lambda=780.241$~nm} is the wavelength  and \mbox{$\Gamma=2\pi\times6.067$~MHz} is the natural linewidth.  The effective detuning of the imaging beam  from  the three possible optical transitions $F=2\rightarrow F'=1,2,3$ weighted by their transition strengths is given by 
\mbox{$\Deff^{-1} = \left( 28/\Delta_{2,3} - 5/\Delta_{2,2} - 3/\Delta_{2,1} \right)/20$.} The far right hand side of Eq.~\ref{eq:theta} gives the rotation angle as a product of the Faraday coefficient $c_\textrm{F}$ and atomic column density $\tilde{n}(x,y)$~\footnote{The Faraday coefficient also depends on the total angular momentum $F$. See \cite{Gajdacz2013} for details on  the calculation of $c_\textrm{F}$ and how it changes for the experimentally relevant values of alkali atom nuclear spin $I=3/2$ and $7/2$.}.

\begin{figure}[tb]
	\centering
	\includegraphics[width=8.3cm]{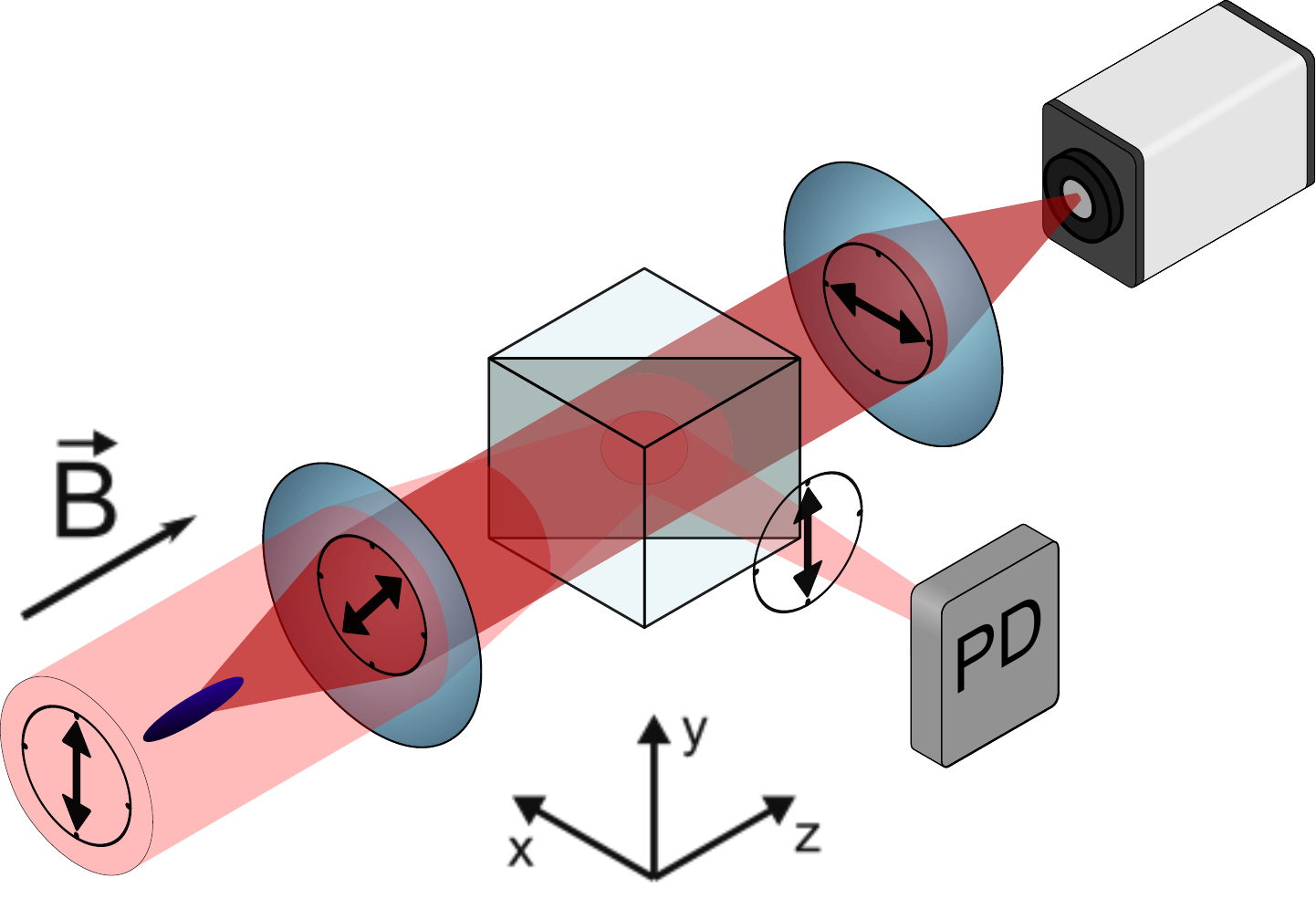}
	\caption{Sketch of the main features of the Faraday imaging setup. The polarisation state of the light is indicated by  double-headed arrows. Through the Faraday effect, the  atomic cloud induces a rotation of the imaging beam's plane of polarisation. A polarizing beam splitter removes the unscattered light from the imaging path, and the transmitted light is imaged on a scientific camera. An additional imaging telescope between the PBS and camera is not shown.}
	\label{fig:Fig1_detectionSetup}
\end{figure}

Figure~\ref{fig:Fig1_detectionSetup} shows the key elements of the  imaging system we use to measure this rotation angle. An off-resonant imaging beam  initially linearly polarised along the $y$ axis propagates along the quantization  axis defined by a bias magnetic field oriented along the $z$ axis. 
The spatial distribution of the atomic cloud is mapped onto the polarisation of the imaging beam through the Faraday effect. A polarising beam splitter (PBS) cube transmits only the rotated light according to the law of Malus, which for an ideal PBS yields $I=I_0\sin^2\theta$, with $I_0$ and $I$ the intensity distribution of the beam before and after the PBS, respectively. The transmitted light is then imaged on a scientific camera. This  realizes a `dark field' dispersive imaging technique in which the non-rotated light is removed from the imaging path \cite{Ketterle}.

\section{Experimental setup}\label{sec:ExpImplementFaradatImg}
The atomic sample  is produced by standard techniques~\cite{Lewandowski2003,Bertelsen2007}.  We first generate a large cloud containing $\sim 10^9$ \Rb atoms  in a vapour pressure magneto-optical trap (MOT), whereupon the atoms are transferred into a magnetic quadrupole trap. The quadrupole trap is   transported on a translation stage to a region of the vacuum chamber that is held at a lower pressure, where the atoms are transferred  to  a Ioffe-Pritchard (IP) magnetic trap \cite{Bertelsen2007,Esslinger1998}. The IP trap has axial and radial trapping frequencies $ \omega_z = 2\pi \times 17.1 \si{.Hz} $ and $ \omega_\rho = 2\pi\times 296 \si{.Hz} $, respectively, with an axial bias field of 330~mG. Forced radio-frequency (RF) evaporative cooling is used to produce ultracold clouds. For the  characterisation of Faraday imaging presented here, we produce clouds in the ${|F=2, m_F = 2 \rangle}$ state containing on average $5.8\times10^6$ atoms at a temperature of $18~\mu$K.

The Faraday imaging light is off-resonant and applied in pulses, with an image  acquired for each imaging pulse. It  is generated by an external cavity diode laser, which is locked with a detuning  \mbox{$ \Delta_{2,3}= 2\pi\times(1200 \pm 1) \si{.MHz} $} to the blue of the $ F=2 \rightarrow F' = 3 $ transition. The detuning is achieved by offset-locking the imaging laser to a master laser, which is locked to a resonance in \Rb by standard saturated absorption spectroscopy~ \cite{[{The offset-lock setup is described in~}][{. This approach provides great flexibility in the choice of detuning, allowing us to vary the detuning by several GHz. In practice, however, we seldom vary the chosen detuning by more than a few hundred MHz, meaning that a simpler setup could also be used, e.g., where the imaging light is derived from quasi-resonant light diffracted by an acousto-optic modulator and/or locked to a nearby transition in another isotope.}] Schuenemann1999}.
At the position of the atoms, the intensity distribution of the imaging light is  Gaussian with a waist of $1.4$~mm, leading to an approximately uniform intensity over the cloud. For the experiments presented here, we use an   intensity of 0.5~mW/cm$^2$. 
The imaging pulses have rectangular temporal envelopes, which are generated by an acousto-optic modulator (AOM).
The optical power of each imaging pulse is monitored on a photodiode (PD) on the reflecting port of the PBS (see Fig.~\ref{fig:Fig1_detectionSetup}) \footnote{Due to the size of the beam relative to the cloud and the modest peak rotation angles, the reduction in photodiode signal due to the atoms is negligible.}. This signal is  used to regulate the amplitude of the pulses through feedback to the AOM to achieve a relative error in the number of photons per pulse of approximately $4\times10^{-3}$.
 The Faraday rotated light that is transmitted through the PBS is imaged on an  Electron Multiplying Charge Coupled Device (EMCCD)
camera. The imaging system has a magnification of 4.85.

\begin{figure}[tbp]
	\centering
		\includegraphics[width=8.3cm]{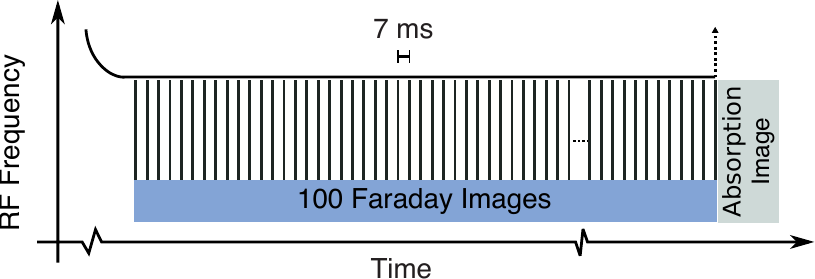}
		\caption{Experimental sequence. An atom cloud held in a magnetic trap is prepared by forced  radio frequency evaporative cooling. The cooling stops when the  cloud contains on average   $N=5.8\times10^{6}$ at a temperature $T=18~\mu$K. We then take 100 Faraday images separated by 7~ms, whereupon the trap is extinguished and a time of flight absorption image is acquired.}
	\label{fig:Figure2}
\end{figure}

Figure~\ref{fig:Figure2} shows the experimental sequence we use to characterise the precision of the Faraday imaging. After initial evaporative cooling, the RF frequency is held constant, producing a trap with an effective depth of 64.3~$\mu$K. A series of 100 Faraday images is taken with \SI{7}{.ms} between the start of each imaging pulse. The  duration of each pulse can be varied to investigate the imaging precision as a function of the total time over which the atoms are probed. After the 100 Faraday images have been taken, the magnetic trap is extinguished and an absorption image is acquired after a time of flight. 

\section{Faraday image processing and accuracy}\label{Image evaluation}
In this section, we describe the  processing steps we perform to obtain high quality Faraday images and to determine the atom number. In principle, 
Eq.~\eqref{eq:theta} provides a link between rotation angles and the column density of the atom cloud. 
However, even for the thermal clouds in this work, density dependent collective light scattering effects lead to  deviations from this light scattering model~\cite{Kaminski2012}. For the imaging parameters used here and consistent with our earlier work~\cite{Gajdacz2013},  the atom number inferred from Faraday imaging is approximately 36\% lower than the one from absorption imaging.

Of course,  absorption imaging itself requires careful calibration to achieve accurate measured atom numbers. We calibrate the absorption imaging by the procedure described in \cite{Reinaudi2007}, whereby a scaling factor for the effective saturation intensity of the imaging transition is empirically determined. An independent method to verify this calibration relies on measuring the scaling of the imaging noise as the atom number is varied \cite{Muessel2013}. 

Such calibration methods invariably make use of  Poissonian statistics in some aspect of the imaging process. 
In our case, we use  the fact that single particle loss from the trap is a Poissonian process \cite{Kampen2007}, such that the atom number variance equals the mean number of atoms; i.e., $\Delta N^2=N$. In recent work, we used Faraday imaging to probe the increase in atom number noise as a function of applied single particle loss~\cite{Gajdacz2016}. Here, we use the single particle loss induced by the imaging itself as a way to link the mean atom number with the variance. Making use of the fact that  imaging accuracy and precision are linked by the requirement of self-consistency  is a generic approach \cite{Hilliard2015}. 

To summarise, we achieve accurate atom numbers from Faraday imaging by calibrating the Faraday images  with  the results from  absorption imaging~\cite{Reinaudi2007,Gajdacz2013}; we then  check this calibration for self-consistency by verifying  the atom noise scaling of the Faraday images.

\begin{figure}[tbp]
	\centering
	\includegraphics[width=8.3cm]{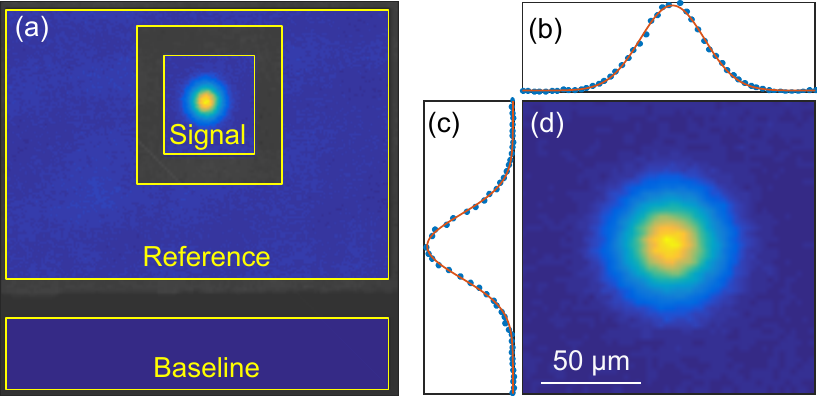}	
	\caption{(a) Full frame of a Faraday Image. The ROIs for signal (innermost square), reference (outermost square, excluding a safety margin) and baseline (bottom square) are indicated. The greyed areas are discarded safety margins. (b,c) Cross-sections of evaluated signal (blue points) with corresponding cross-sections of the fit (red line). (d) Evaluated signal from a single image.}
	\label{fig:Fig2_ImageEvaluation}
\end{figure}

The  light detected on the EMCCD camera scales with the Faraday `signal' defined as 
\begin{equation}\label{eq:SignalDefinition}
S \equiv \frac{I(\theta)}{I_0}=\sin^2(\theta).
\end{equation} 
Due to the finite extinction ratio of the PBS, (`cube suppression') $ \CS = \num{1.5e-3} $, a small fraction of the non-rotated light leaks through the cube. This leads to an intensity after the PBS given by: \mbox{$I(\theta)=I_0[\sin^2(\theta)+\CS\cos^2(\theta)][1+\CS]^{-1}$}~\cite{Gajdacz2013}. Experimentally,  we  calculate the signal in a region-of-interest (ROI) that encompasses the cloud as 
\begin{equation}\label{eq:SignalDefinitionExpt}
S = \left(\frac{I(\theta)}{I_\mathrm{ref}} - 1\right)\left(\frac{\CS}{1-\CS}\right),
\end{equation}
where $ I_\mathrm{ref} \equiv \CS\cdot I_0$ is the non-rotated light leaking through the cube; i.e., we use $ I_\mathrm{ref}$ as a measure of $I_0$ in a given image so that we avoid having to take an additional picture without the atoms.
We obtain $ I_\mathrm{ref} $ by taking the mean pixel count in a ROI outside the area covered by the atomic cloud. Both  $ I(\theta) $ and $ I_\mathrm{ref} $ are corrected for the baseline level of the camera, which is the mean pixel count evaluated in a masked area of the camera. Figure~\ref{fig:Fig2_ImageEvaluation}(a) illustrates the regions of interest where we measure the signal intensity $ I(\theta) $, reference intensity $ I_\mathrm{ref} $ and baseline. Figure~\ref{fig:Fig2_ImageEvaluation}(d) shows the resulting evaluated signal in its ROI.

To evaluate the Faraday images, we  can  work directly  with the signal $ S $ or  evaluate the distribution of rotation angles $ \theta $ induced by the cloud. 

If we work with the rotation angles,  the cloud parameters are obtained by   fitting the  signal with: $S_\textrm{G}\sim\sin^2\left(\theta_\mathrm{G}\right)$, where $\theta_\mathrm{G}$ is a 2D Gaussian distribution. This   corresponds to the in-trap column density distribution $\tilde{n}$ for  a thermal cloud characterised by an atom number $ N $ and  temperature $ T $ ~\cite{Pethick2008}, scaled by the Faraday coefficient $c_\textrm{F}$ (see Eq.~\eqref{eq:theta}). Figures~\ref{fig:Fig2_ImageEvaluation}(b) and (c)  show cross-sections of   $ S $ and this fit. The area integral of the fitted  distribution $\theta_\mathrm{G}$ thus  provides a measure of the total atom number, which we denote $ \angsum $ \footnote{The Faraday rotation angles can also be reconstructed from the evaluated signal through \mbox{$ \theta = \arcsin(\sqrt{S})$} \cite{2016arXiv160702934B}, but this approach can be problematic.
 First, since the baseline is subtracted, the pixel values where the signal is low can become negative, which complicates the  evaluation of  the square root. Secondly, the steep derivative of the square root for small arguments amplifies the noise in pixels with low signal.}.

\begin{figure}[tbp]
	\centering
	\includegraphics[width=8.3cm]{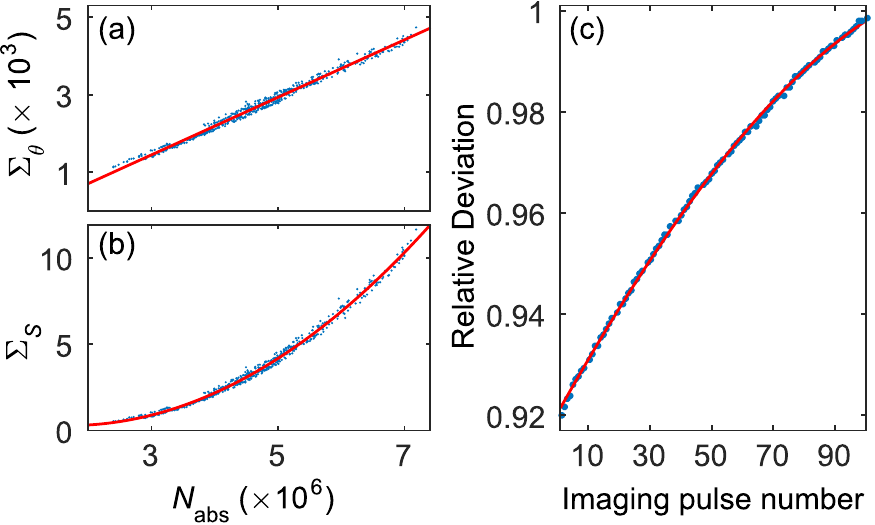}
	\caption{Accurate atom numbers can be obtained through cross-calibration of Faraday imaging with absorption imaging. (a) Angle sum as a function of the atom number obtained from absorption images (blue points) and fitted first order polynomial (red line). (b) Signal sum as a function of the atom number obtained from absorption images (blue points) and fitted second order polynomial (red line). (c) Ratio between atom number obtained from the angle sum using the calibration in (a) and the atom number obtained from the signal sum using the calibration in (b) as a function of imaging pulse number. The ratio is 1 for the 100'th image, since both the signal sum and the angle sum have been calibrated with the atom number obtained from an absorption image taken directly after the final Faraday image.}
	\label{fig:Calibration}
\end{figure}

Figure~\ref{fig:Calibration}(a) shows $ \angsum $  in the 100'th Faraday image of each experimental run as a function of the  atom number extracted from the subsequent absorption image. The figure includes several data sets corresponding to different values of imaging pulse duration, taken over several days. The range in atom numbers is achieved through the atom loss induced by the imaging light and through the  spread in initial  atom number due to technical   fluctuations in the  experiment; the standard deviation in initial atom number over all data sets is 12.5\%.
The data has been fitted by a linear function: $ \angsum = a_1 (N - a_2) $.
Due to the finite extinction ratio of the PBS, there is a lower bound for detection of small Faraday rotation angles, which is modelled by the offset $ a_2 $. 
The calibration allows us to convert the angle sum  obtained for each  image into an atom number.

An alternative approach is to work only with the total signal sum $ \sigsum = \sum S_{ij} $, where the sum runs over all pixels in the signal ROI. Since $\sigsum$ is a scalar function of $N$ and $T$, this processing method does not provide independent measures of the cloud's parameters. 
In practice, however, $N$ and $T$ are  highly correlated in an experimental run because the evaporative cooling relies on elastic collisions between  atoms \cite{Ketterle1996}. This means that $\sigsum$ can be calibrated through its correlation with $N$ and $T$ obtained from absorption imaging ~\cite{Gajdacz2016}, and the calibration in Fig.~\ref{fig:Calibration}(a), which is linear in the atom number.
Despite these complications,  we  show in Sec.~\ref{sect:Precision} that working directly with the signal sum $ \sigsum $ yields superior imaging precision over the angle sum $ \angsum $ because we avoid the noise induced by model fitting. Additionally,
this simple analysis is well-suited for implementation on a field programmable gate array (FPGA), which has allowed us to obtain real-time information on the cloud properties during individual experimental runs \cite{Gajdacz2016}.

   Figure~\ref{fig:Calibration}(b) shows the signal sum $\sigsum$  in the 100'th Faraday image as a function of the  atom number extracted from the subsequent absorption image, for the same datasets as in Fig.~\ref{fig:Calibration}(a).
   The data has been fitted by a quadratic function, which is motivated by the scaling of the signal sum in the limit of small Faraday rotation angles: for the thermal clouds employed in this work and our choice of detuning, the  largest rotation angle (at the centre of the cloud) is small, i.e., $ \theta< 15^\circ $, such that the dependence of  $ \sigsum $ on atom number and temperature can be approximated by the first order series expansion of Eq.~\eqref{eq:SignalDefinition}: $S \approx \theta^2=(c_\textrm{F}\tilde{n})^2$. Inserting a 2D Gaussian distribution for the column density yields the small angle dependence of the signal sum on the atom number and temperature: $\sigsum \propto {N^2}/{T}$. The exact form of the quadratic fit function contains additional parameters used to model technical details in the imaging. The function is given by: $ \sigsum = b_1(N - b_2)^2 + b_3 $.
Here $ b_2 $ again models a lower bound in the detection of small rotation angles. The parameter $ b_3 $ models imaging beam inhomogeneity, which gives rise to an offset in the detected signal sum.

This approach is based on the  simplifying approximation that the temperature of the sample is unchanged due to the imaging. For the range of imaging pulse durations we investigate here, we  observe that the cloud temperature variation is less than $1.5 \%$, as determined from the absorption imaging. In Appendix~\ref{Atomloss}, we show that the heating induced by photon scattering from the imaging beam is negligible compared to the temperature of the sample, and that the dominant  loss mechanism is spontaneous decay into  states that are not trapped by the magnetic potential.

While the imaging does not influence the temperature of the sample, the cloud cools down during imaging due to free evaporation. Estimating the atom number based on the signal sum calibration in Fig.~\ref{fig:Calibration}(b) alone,   leads to a systematic underestimation of the atom number in the earlier images because the cloud is hotter there than in later images. This occurs due to the coupling of $ N $ and $ T $ which may be approximately understood from the small angle dependence of the signal sum.

  Figure~\ref{fig:Calibration}(c) shows the ratio of the atom number obtained  from the signal sum using the fitted quadratic calibration function in Fig.~\ref{fig:Calibration}(b) and the atom number evaluated 
  from the angle sum using the fitted linear calibration function in figure~\ref{fig:Calibration}(a). For the first image, the signal sum underestimates the atom number by $ 8 \% $ decreasing to zero for the 100'th image. The atom number obtained from $\sigsum$ in the final image is accurate due to the back-to-back comparison with absorption imaging. The data has been fitted by a quadratic function, allowing us to correct for the underestimation and thus obtain an accurate atom number for each image from the signal sum alone.

To summarise this section, we  employ two main methods to obtain simple measures of the atom number from the Faraday images. In the first method, we fit a model of the signal transmitted by the PBS, from which we obtain $\angsum$, a quantity proportional to the atom number. In the second, we simply sum the pixels in a region of $S$ that encompasses the cloud to obtain $\sigsum$, which in the small rotation angle limit scales as $N^2/T$. Density dependent light scattering effects lead to deviations from the simple proportional dependence between column density and rotation angles in Eq.~\eqref{eq:theta}, so to obtain accurate atom numbers, we calibrate  $\angsum$ and $\sigsum$ in the final Faraday image with the atom number obtained from absorption imaging. The nonlinear scaling of $\sigsum$ with $N$ and $T$ necessitates an additional calibration of $\sigsum$ with $\angsum$ to obtain accurate atom numbers.   
In the following section,  
we first  characterise the noise properties of the Faraday  imaging method for the case of single  images. Subsequently, we investigate how the noise properties change when we average over several images.

\section{Faraday Imaging Precision}\label{sect:Precision}
\begin{figure}[tbp]
	\centering
	\includegraphics[width=8.3cm]{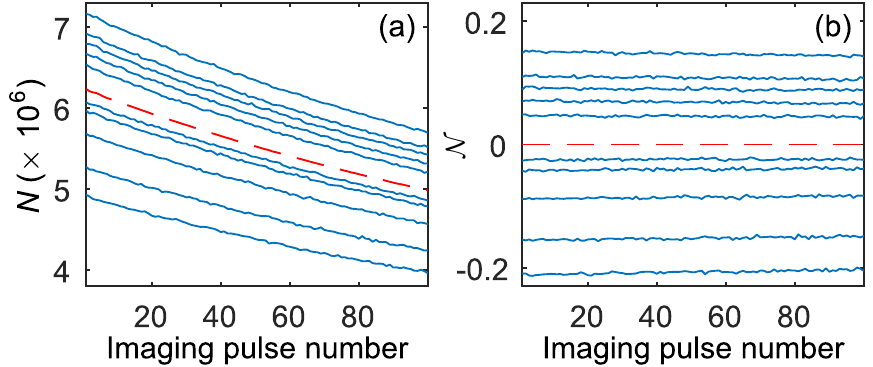}
		\caption{Evolution of the atom number during Faraday imaging. (a) Representative traces for  a pulse duration of \SI{0.66}{ms} showing the decay of atoms during imaging. Red dashed line indicates the mean atom number over the entire dataset. (b) Traces showing relative deviation $\mathcal{N}$ from the mean atom number.}
	\label{fig:traces}
\end{figure}
To obtain the best possible precision in any detection scheme, care has to be taken to minimise any sources of technical noise, such that one can approach the limits set by the inherent fundamental noise sources. In our imaging method, the fundamental noise sources are photon shot noise and the stochastic uncertainty of the number of lost atoms induced by the imaging light. {In addition}, the images suffer from  technical noise in the camera and the imaging beam.  The technical noise contributions from the imaging beam are instability in the laser  frequency, instability in the amplitude of the imaging pulses  and mechanical vibrations of the optical components in the imaging beam path. Of these noise contributions, we judge the uncertainty in the detuning to be the dominant technical noise source, given that  the imaging pulses are actively stabilised in amplitude (see Section~\ref{sec:ExpImplementFaradatImg}). 
The noise sources in the camera are readout noise, clock induced charges, thermal dark counts and spurious charges. With modern  scientific cameras, the readout noise is typically the dominant noise source. In our method, the read-out noise is most deleterious  when determining $I_\mathrm{ref}$ where the integrated number of photons in an image is small and in the wings of the cloud, where the rotation angles are small. To overcome the readout noise, we  take advantage of the camera's electron multiplying (EM) feature to amplify the detected signal above the readout noise. The EM amplifier consists of a multi-stage avalanche-type amplification, and leads to a multiplicative noise term increasing the photon shot noise by a factor approaching $ \sqrt{2}  $ \cite{Robbins2003,Hynecek2003}. By using large ROIs to estimate $I_\mathrm{ref}$ and the baseline, we ensure that the imaging precision is limited by the noise  in the signal region. The application of EM-gain reduces the size requirement of the reference ROI.

\subsection{Single image detection}\label{sec:SingleImgCharac}
Figure \ref{fig:traces}(a) shows the atom number measured  with an imaging pulse duration of 0.66~ms as a function of imaging pulse number. The atom number has been obtained from the signal sum using the fitted  quadratic  function in Fig.~\ref{fig:Calibration}(b), and corrected for the variation in temperature using the fitted quadratic correction in Fig.~\ref{fig:Calibration}(c).  
The fluctuations of the atom number about the mean decay come from the light shot noise, the stochastic noise arising from atom loss, and potential technical noise. Given that the mean atom loss is a deterministic process, we can remove the decay by subtracting  or dividing $N$ by the appropriate exponential curve. For simplicity, we normalize the atom number  by its mean over several experimental runs and shift it to be centered on zero, yielding the `normalised atom number'
\begin{equation}\label{eq:Error}
\Nnorm_{i,j} = \frac{N_{i,j}}{\mean{N_{i,j}}_\mathrm{runs}} - 1,
\end{equation}
where $ i $ denotes the run number and $ j $ denotes the image number in a given run. $ \mean{N_{i,j}}_\mathrm{runs} $ is the mean atom number in the $ j $'th image, averaged over all runs in the  dataset. 
  Figure~\ref{fig:traces}(b)  shows the atom number  traces from \ref{fig:traces}(a) in this normalised form, where the fluctuations about each  trace give the relative uncertainty in $N$.

\begin{figure}[tbp]
	\centering
	\includegraphics[width=8.3cm]{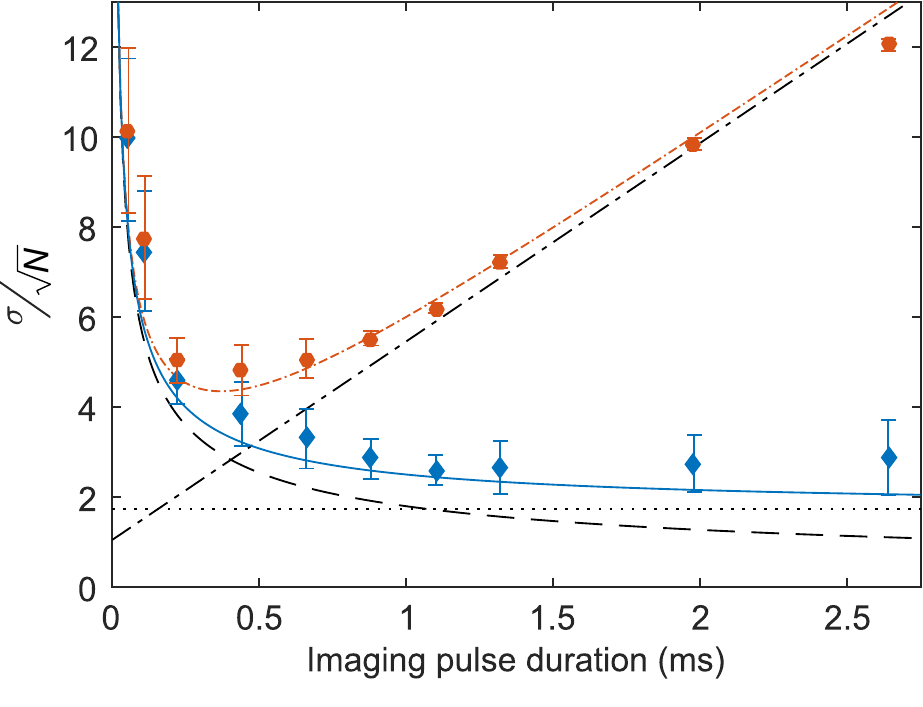}
		\caption{Relative two-sample deviation of atom number (red circles) and normalised atom number (blue diamonds) as a function of the pulse duration. The photon shot noise (dashed line) is dominant for short pulse durations. As the pulse duration is increased, the mean atom decay contribution (dash-dot line) begins to dominate $\sigma_N$. By evaluating the normalised atom number, the mean decay can be eliminated, such that   $\sigma_\mathcal{N}$ continues to decrease as the imaging pulse duration is increased; this noise  approaches a constant offset (dotted line) for large $t$, which we attribute to technical noise in the imaging light. The total noise model (blue line) for the normalised atom number is comprised of the photon shot noise and technical noise contributions added in quadrature. The corresponding curve (red dash-dash-dot line) for the atom number is comprised of the photon shot noise, technical noise and mean atom loss noise contributions added in quadrature.
 The  decay per image is so small that the stochastic atom loss noise is negligible. Error bars represent 1-$ \sigma $ (standard deviation) statistical uncertainties.}
	\label{fig:SingleImagePrecision}
\end{figure}

One of the most attractive experimental features of taking multiple images of the same cloud is the ability to use the 
statistics of the imaging series to provide an \textit{in situ} measure of the imaging precision.
To characterise the measurement noise, we evaluate the two-sample variance for all pairs of images along the individual traces. For the atom number, this is given by
\begin{equation}\label{eq:twoSampleVariance}
\sigma_{N}^2 = \frac{1}{2}\mean{(N_{j+1} - N_j)^2}.
\end{equation}
The two-sample variance can be evaluated for  the atom number and normalised atom number traces shown in Figs.~\ref{fig:traces}(a) and (b). The two-sample variance is insensitive to sufficiently slow trends or drifts in the measurements, but captures the relevant noise. 

Figure~\ref{fig:SingleImagePrecision} shows the relative two-sample deviation,   for  the atom number and  normalised atom number as a function of imaging pulse duration $t$. The two-sample deviation is defined as the square root of the two-sample variance, and it is plotted relative to $\sqrt{N}$; with this choice of scaling, the Poissonian noise level is achieved at $\sigma/\sqrt{N}=1$. 
Initially, the two-sample deviation  of the  atom number  decreases as the imaging pulse duration is increased, but at 0.5~ms it reaches a minimum and then increases approximately linearly with~$t$. The two-sample deviation  of $\Nnorm$  shows the same initial behavior, but does not  increase for the range of $t$ we consider because the mean atom loss contribution to the noise has been removed through the normalisation.
This shows that in a single Faraday image, the imaging noise lies within a factor of 3 of the Poissonian noise level. 
To gain insight into the contributions of the fundamental noise sources, we evaluate the photon shot noise and the stochastic noise due to loss of atoms: the detection of  photons and the loss of atoms are examples of single particle events leading to Poissonian statistics.

The photon shot noise can be evaluated directly from the Faraday images.  The  number of photo-electrons  generated when an incident photon hits the CCD is  Poisson distributed; i.e.,  the variance in the number of generated photo-electrons is equal to the mean number $ N_\mathrm{el} $.  By error propagation, the uncertainty in the number of photo-electrons leads to an uncertainty in the detected signal given by \cite{Gajdacz2013}: $\sigma^2_{\sigsum} = {{\sigsum}/{N_\mathrm{el,0}}},$
where $ N_\mathrm{el,0} $ is the number of electrons per pixel we would detect if the cloud and PBS were absent. We  evaluate this from the number of counts in the reference area by taking the measured camera gain and  cube suppression into account. Hence, the  uncertainty in atom number arising from the photon shot noise can be calculated as 
$\sigma^2_\mathrm{psn} =( \partial N/\partial \Sigma_S)^2 \sigma^2_{\sigsum}\equiv \beta^2  \sigma^2_{\sigsum} $, 
where the error propagation coefficient is 
 obtained from the calibrations in Figures.~\ref{fig:Calibration}(b) and (c). It is evident from Fig.~\ref{fig:SingleImagePrecision}, that the photon shot noise is the dominant noise source for short pulses, but decreases as $ 1/\sqrt{t} $ as the pulse duration  is increased.

Atom loss leads to an increase in the two-sample variance 
 due to both the stochastic nature of the discrete particle loss and through the decay of the mean atom number.
  Assuming that only single-particle loss occurs, the two-sample variance due to loss is given by \cite{Hume2013}
\begin{equation}\label{eq:stochLoss}
\sigma_\mathrm{loss}^2 = \frac{N_0}{2\tau}t + \frac{N_0^2}{2\tau^2}t^2,
\end{equation}
where $ N_0 $ is the atom number at the start of the imaging and $ \tau $ is the  lifetime during imaging, such that the mean atom number is given by $N=N_0\exp(-t/\tau)$. The first term in Eq.~\eqref{eq:stochLoss} describes  stochastic atom loss noise and the second describes the effect of  mean decay on the two-sample variance.
Atom loss due to residual thermalization and background losses gives rise to a small extra contribution of atom loss noise, leading to an offset in the mean atom loss noise curve  in Fig.~\ref{fig:SingleImagePrecision}. The lifetime $\tau$ has been estimated by fitting the atom number traces. The second term can be eliminated by removing the mean decay from the traces as we do for the normalised atom number $\Nnorm$. For all the pulse durations considered here, the stochastic noise in a single image is negligible.

Figure~\ref{fig:SingleImagePrecision} shows that the photon shot noise $ \sigma_\mathrm{psn} $ and the atom loss noise $ \sigma_\mathrm{loss} $ plus a  constant technical noise contribution $ \sigma_\mathrm{tech} $ account extremely well for the observed noise in single images. We emphasize that the photon shot noise and atom loss noise models include no fitted parameters: they are evaluated directly from the Faraday image traces. The dominant source of  the technical noise   is likely to 
be  the frequency stability of the detection laser: the laser has $\sim$\SI{1}{\mega\hertz} linewidth in a 5 ms integration time, corresponding to $ \sim \num{1e-3} $ relative uncertainty in the detuning. By error propagation using Eq.~\eqref{eq:theta}, this  corresponds to $ \sim \num{8e-4} $ relative uncertainty in the detected number of atoms; note that $ \num{8e-4}\cdot\sqrt{N}\approx 1.9 $ for $N=5\times10^{6}$, in agreement with Fig.~\ref{fig:SingleImagePrecision}. The integration time for the linewidth measurement is similar to the time between consecutive images, and  we therefore expect this technical noise contribution to be independent of the pulse duration. The technical fluctuations in the imaging pulse amplitude at the $4\times10^{-3}$ level (see Section~\ref{sec:ExpImplementFaradatImg}) can be regarded as a next order effect, in that these fluctuations are present in both the signal and reference ROIs and hence accounted for in Eq.~\eqref{eq:SignalDefinitionExpt}. The  fluctuations in the imaging pulse amplitude  manifest themselves  instead as small variations in the mean atom decay rate for each imaging pulse.   

In summary, we have investigated the noise properties in the individual Faraday images and characterised this noise with an imaging model that includes photon shot noise, atom loss noise and technical noise associated primarily with the imaging laser linewidth. The noise arising from stochastic atom loss is negligible for our imaging parameters, such that only the mean atom loss noise is manifest in the two-sample variance. The optimal imaging pulse duration of approximately 0.45~ms corresponds to the point where the photon shot noise and mean atom loss terms are equal, as one would expect when adding two quantities in quadrature. For this optimal setting, we achieve a single image noise level that is approximately five times the atom shot noise level. If we instead work with the normalised atom number $\mathcal{N}$, such that the mean atom loss noise contribution is absent,  the imaging noise is reduced by 20\% and approaches the technical noise floor for longer imaging pulses.

\subsection{Multi-image detection}\label{sec:MultiImgCharac}

While  the single image noise  decreases as the pulse duration is increased when the data is corrected for the mean atom loss,   the pulse duration is in practice limited by the dynamic range of the camera. The problem of finite dynamic range is exacerbated in our experimental realization by the fact that significant EM-gain is necessary to render the readout noise negligible when imaging small rotation angles. To artificially increase the dynamic range of the camera, we can exploit the `non-destructive' nature of our detection, and take multiple images and average over the result. For uncorrelated noise, the standard error of the mean decreases with the number of images $ n $ as $ 1/\sqrt{n} $. This is the case for the shot noise and a fair assumption for the technical noise $ \sigma_\mathrm{loss} $. It is not the case, however, for  stochastic atom loss noise: this noise is  correlated because the width of the atom number distribution reflects the random walk in atom number during all previous images.

To obtain experimental measures of the detection precision when averaging over several images, we break the data  used for the single image characterisation into two halves: the first set contains images 1:50 and the second set contains images 51:100. The number of images over which we average  can be varied for $1\leq n\leq 50$.
The  normalised atom number averaged over images \mbox{$[50-(n+1),50]$} is denoted by $\Nnormone ^{n}$, and over images \mbox{$[51,51+(n-1)]$} by $\Nnormtwo ^{n}$. We focus on an intermediate imaging pulse duration, $t=0.88$~ms. 

\begin{figure}[btp]
	\centering
	\includegraphics[width=8.3cm]{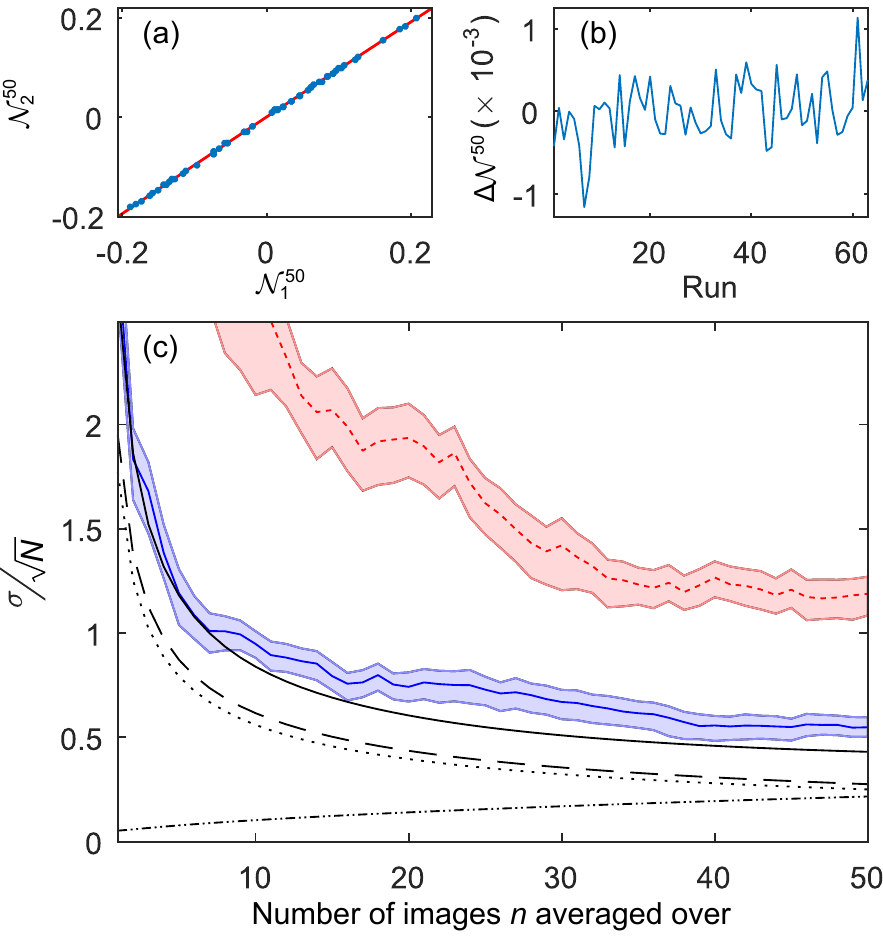}
	\caption{Characterisation of imaging noise when averaging over several images. (a) Correlation of normalised atom number averaged over 50 images in the first and second image set. Line is a quadratic fit to the data. (b) Deviation of the data  from the fit in (a), plotted as a function of experimental run number. (c) Experimental imaging noise and noise models as a function of the number of images over which the average is performed. The blue curve shows the measured experimental noise. The red curve (fine dashed) shows the experimental imaging noise when the atom number is  calculated by directly fitting the Faraday images. The  shadings represent  one standard deviation error bars obtained by bootstrap-resampling the data with replacement. 
	Imaging model curves correspond to stochastic atom loss noise (dash-dot-dot), technical noise (dotted), photon shot noise (dashed) and total noise model  (solid).} 
	\label{fig:MultiImDetection}
\end{figure}

Figure~\ref{fig:MultiImDetection}(a) shows $\Nnormtwo ^{50}$ as a function of $\Nnormone ^{50}$, i.e., we compare the results of the first and second image set where the average is performed over all 50 images. The 10\% standard deviation   in the initial atom number  for this data set  (determined from the  first Faraday image in each run)   is due to fluctuations in the experimental  apparatus.
It is evident that the normalised atom numbers in the first and second image sets are highly correlated. To extract the imaging noise, we fit a quadratic function  to this data and evaluate the deviations  from the fit, which we denote $\Delta \Nnorm ^{50}$. A quadratic fit is used because  the atom number extracted from the signal   retains  vestiges of the temperature dependence of $\sigsum$ (see Section~\ref{Image evaluation}). However, this systematic  calibration effect is extremely small,  as is clear from the data, and a linear fit  yields comparable results.

The need for the two-sample variance in order to characterise the imaging noise while rejecting drifts in the experimental apparatus is illustrated in Fig.~\ref{fig:MultiImDetection}(b), which shows $\Delta \Nnorm ^{50}$ as a function of  run number. These deviations 
exhibit the noise we wish to determine, on top of a slight upward trend. This trend is caused by  environmental drift that is slow compared to the $\sim 1$ minute duration of an experimental run. The primary cause of drift in the experiment is currently $\sim 0.1^\circ$C variations in the  water used to cool the magnetic trap: as the temperature of the cooling water drifts, the magnetic trap coils expand and contract slightly,  leading to small drifts in  the trapping parameters. In particular, the magnetic trap bottom drifts with respect to the RF knife that determines the height of the trapping potential, leading to slight changes in the atom loss rate due to free evaporation. 

Figure~\ref{fig:MultiImDetection}(c) shows the measured imaging noise  as a function of the number of images included in the average. 
The blue line shows the measured imaging noise, evaluated by  calculating the two-sample variance on $\Delta \Nnorm ^{n}$  ordered by run number, for  $1\leq n\leq 50$. The shaded region denotes the uncertainty in the measured noise calculated by bootstrapping~\cite{Efron1979}.  As expected, the noise is seen to decrease with the number of images $n$ used in the average.

The measured imaging noise is well described by a noise model comprised of contributions from the photon shot noise, stochastic atom loss noise and technical noise, evaluated to include the effect of averaging. For the photon shot noise and technical noise, we use the results from the single image precision, scaled  by $1/n$. A noise contribution  $\sigma_\mathrm{stoch}$ to describe the effect of   stochastic atom loss when averaging over several images is derived in Appendix~\ref{Atomnoisemodel}.
    The  noise model with  these contributions is given by \mbox{$\sigma^2 = { {\sigma_\mathrm{psn}^2}/{n} +{ \sigma_\mathrm{tech}^2}/{n} + \sigma^2_\mathrm{stoch}}$.}
The agreement between the noise model and the data is excellent, given that the model contains no free parameters: it  is  calculated directly from the statistics of the experimental data and an independent estimate of the imaging light frequency stability.
The experimental noise and model  plotted in Fig.\ref{fig:MultiImDetection}(c)  are again shown as the square root of the two-sample variance, scaled by the mean atom number in images 51 and 52. With this presentation, it is evident that we achieve sub-atom shot noise  detection for averages over 10 or more  images. Additionally, by reducing the imaging laser linewidth to a realistic value of $\sim100$~kHz, the technical  contribution to the imaging noise could be rendered negligible and we could  attain an imaging noise level of  $\sigma/\sqrt{N}\approx0.4$.

The same imaging performance analysis can be performed using $\angsum$, whereby  the atom numbers are extracted by directly fitting the signal distributions in the Faraday images. The result is shown as the red (fine  dashed) line and shaded region in Fig.~\ref{fig:MultiImDetection}(c). We also obtain excellent precision with this method, but as expected the fitting routine introduces extra uncertainty, and we do not achieve sub-shot noise limited detection. Interestingly, it is beneficial to average over more images when using this approach. Finally,  we note that  this approach, up to a scaling factor,  allows the total atom number to be evaluated with high precision even in cases where cross-calibration with absorption images is not possible.

\section{Conclusion}\label{sect:Conclusion}
In this work, we have demonstrated that a simple dispersive imaging technique  can  measure the atom number in a large, ultracold  cloud with a precision more than a factor of two below the atom shot noise level. The minimally destructive character of the technique allows us to take multiple images of the same cloud. The benefit of this  is two-fold. First,  by averaging over several images, we improve the precision of the atom number measurement. Second, by analysing the statistics of multiple measurements, we obtain an \textit{in situ} estimate of the measurement precision. 
To understand the achievable precision of the  method, we have developed  an imaging noise model. This model has two fundamental contributions: the photon shot noise on the detected light and the noise associated with the  single atom loss that is induced by the imaging. Through our analysis, we obtain the size of these two contributions directly from the images, in contrast to previous work in which they were estimated from model fitting \cite{Gajdacz2016}. We find that an additional technical noise contribution is necessary to quantitatively describe the imaging noise. We ascribe the origin of this noise to the finite linewidth of the imaging laser, and will in future work render this contribution negligible with an improved laser system.

By comparing  the atom number from Faraday imaging with that obtained from independently calibrated absorption imaging, we  achieve accurate as well as precise atom counting. The accuracy of this approach is supported by the quantitative agreement of the  noise model with the observed imaging noise: the  atom loss contribution to the noise model reflects the Poissonian statistics of the atom number distribution, thereby providing a link between the mean and the variance of of atom number.

\begin{acknowledgments}
We acknowledge support from the Danish Council for Independent Research, the European Research Council, and the Lundbeck Foundation. C.K. acknowledges support from  the Centre for Quantum Engineering and Space-Time Research (QUEST) and  the Deutsche Forschungsgemeinschaft (Research Training Group 1729 and CRC 1227).
\end{acknowledgments}

\appendix
\section{Atom loss during imaging}\label{Atomloss}
The observed loss of atoms due to imaging is caused by spontaneous photon scattering. There are two main processes that arise from this scattering to consider: recoil heating inducing loss due to finite trap depth, and decay into untrapped states.
To determine the size of these contributions, we first estimate the number of scattered photons per atom. To simplify the analysis, we focus on an intermediate imaging pulse duration $t=0.66$~ms. This pulse duration leads to an $11\%$ loss of atoms over 50 images (see Fig.~\ref{fig:traces}(a)), . 

Since the bias magnetic field is oriented along the probing direction, the linearly polarized imaging  light is most easily described as an equal superposition of $\sigma_+$ and $\sigma_-$ circular polarizations. The scattering of photons on the $\sigma_+$ transition does not lead to a change in the atom's internal state, since by selection rules the atom must decay back to $|F=2, m_F=2\rangle$.
The scattering rate on this cycling transition is given by
\begin{equation}
R_\mathrm{scat,+} = \frac{\Gamma}{2} \frac{I_+/I_\mathrm{sat}}{1 + I_+/I_\mathrm{sat} + 4(\Delta_{2,3}/\Gamma)^2 },
\end{equation}
where the saturation intensity of the transition is \mbox{$I_\mathrm{sat} = 1.669~\mathrm{mW/cm^2}$}, and the other quantities were introduced in Sections~\ref{sec:DFFI} and \ref{sec:ExpImplementFaradatImg}.
  The light intensity of the $\sigma_+$ polarized light is $I_+ = I_0 /2$, where the total light intensity typically used in our experiment is $I_0 = 0.5\,\mathrm{mW/cm^2}$. 
For 50 images, the number of scattered $\sigma_+$ photons per atom is
\begin{equation}
N_\mathrm{ph,+} = 50\cdot t \cdot R_\mathrm{scat,+} \approx 0.59.
\label{eq:sigPlusScat}
\end{equation}  

To estimate the number of photons scattered by the $\sigma_-$ light, and with a view to calculating the probability for an atom to be optically pumped to an untrapped state, we must calculate the scattering rate into the three possible excited states:
 $|F^\prime = j, m_F = 1 \rangle$, $j = 1,2,3$. Depending on the excited state, spontaneous decay can occur into  the $F=1$ and $F=2$ ground state manifolds. The decay probability $P_{j,k}$ for the $k = 1,...,5$ final states is proportional to the square of the  matrix element for the transition~\cite{SteckRB87}
\begin{equation} P_{j,k}\propto
\left(
\begin{tabular}{c c | c  c  c}
	& & & $F^\prime = j$ &  \\
	$k$ & & 1 & 2 & 3 \\
	\midrule
	1 & $|F=1, m_F = 0\rangle$ & 5/24 & 1/8  & 0 \\
	2 & $|F=1, m_F = 1\rangle$ & 5/24 & 1/8  & 0 \\
	3 & $|F=2, m_F = 0\rangle$ & 1/120 & 1/8 & 1/5 \\
	4 & $|F=2, m_F = 1\rangle$ & 1/40 & 1/24  & 4/15 \\
	5 & $|F=2, m_F = 2\rangle$ & 1/20 & 1/12  & 1/30 \\
\end{tabular}\right).\label{Table:Matrix}
\end{equation}
In this notation, the strength of the $\sigma_+$ transition is $P_+ \propto 1/2$.
To estimate the number of photons scattered on the individual $\sigma_-$ transitions, $N_\mathrm{ph,j}$,
we also need to take into account the detuning. In the off-resonant limit ($[\Delta/\Gamma]^2 \gg 1 $), we can approximate the number of scattered photons as
\begin{equation}
N_\mathrm{ph,j} \approx N_\mathrm{ph,+} 
\left(\frac{\Delta_{2,3}}{\Delta_{2,j}}\right)^2 \frac{P_{j,5}}{P_+}.
\end{equation}
The total number of photons scattered on the $\sigma_-$ transition is then
\begin{equation}
N_\mathrm{ph,-} = \sum_{j=1}^3 N_\mathrm{ph,j} \approx 
0.032 + 0.066 + 0.039 =  0.137,
\end{equation} 
which is about factor of four lower than $N_\mathrm{ph,+}$. 

Thus, the total number of  photons scattered during an imaging series comprised of 50 pulses with $t=0.66$~ms is given by $N_\mathrm{ph,+}+N_\mathrm{ph,-}\approx 0.7$.
Assuming that each (absorption and emission) scattering event heats  an atom by  the recoil temperature $T_R \approx 0.36\,\mathrm{\mu K}$, the increase in  temperature  is small compared to the temperature of the  cloud: with the initial temperature of the cloud at $ 18~\mathrm{\mu K}$, this corresponds to a temperature increase  of about 1.5\%. 
 The loss of atoms through recoil heating and evaporation
is therefore much smaller  than the observed $11~\%$ atom loss in the experiment.

\begin{figure}[t]
	\centering
	\includegraphics[width=8.3cm]{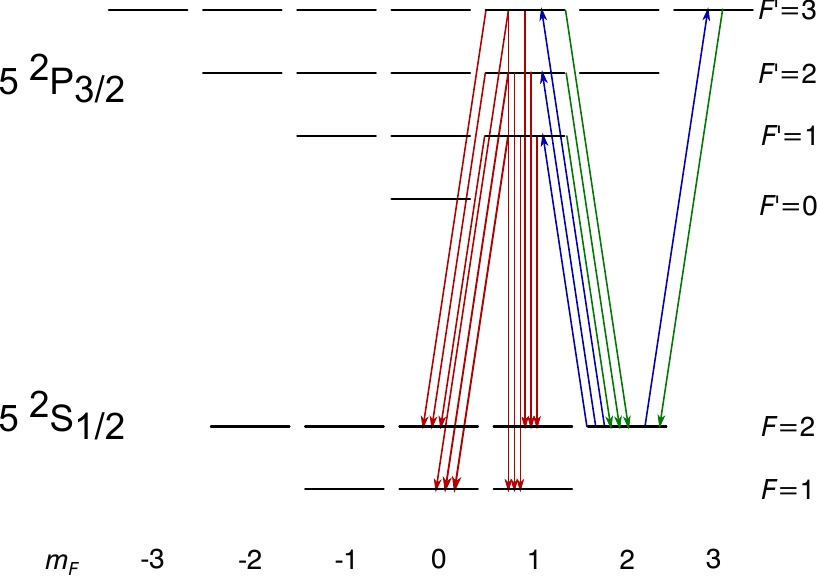}
	\caption{Simplifed level diagram of \Rb to illustrate state transfer due to optical pumping by the imaging beam from 
	\mbox{$|F=2,m_F=2\rangle$.}}
	\label{fig:Figure4}
\end{figure}
We will now estimate the loss due to state transfer. As shown in ~\eqref{Table:Matrix}, the $\sigma_-$  light can optically pump the atoms into three possible excited states.
Decay into any state but $|F=2, m_F=1\rangle$ and $|F=2, m_F=2\rangle$ leads to the loss of an atom from the magnetic trap, whereby the number of scattering events leading to  loss can be estimated as
\begin{equation} 
N_\mathrm{ph,-,loss} \equiv  \sum_{j=1}^3 N_\mathrm{ph,j} 
\frac{ \sum_{k=1}^3 P_{j,k} }{\sum_{k=1}^5 P_{j,k}} \approx 0.092.
\end{equation} 
This result is consistent with the observed loss of $\sim 11\%$ when one takes into account the additional residual background loss of approximately 2\% in the imaging time.

\section{Stochastic noise from atom loss when averaging over several images}\label{Atomnoisemodel}

To see how the stochastic noise grows as the number of images increases, we assume for the moment that the images are only susceptible to the stochastic loss of atoms, and no other detection noise influences the precision. After each image, a certain mean fraction of atoms remains, characterised by the survival probability $ p \equiv N/N_0= \exp(-t/\tau)$. The mean atom number $ \mean{N_k} $ in the $ k $'th image is
\begin{equation}\label{eq:MeanAtomNumber}
\mean{N_k} = \mean{N_{k-1}}p = N_0 p^k.
\end{equation}
Due to our assumption that stochastic atom loss  is the only noise source, $ N_0 $ denotes the initial `true' atom number. The loss of atoms leads to a stochastic step $ \Delta N_k $ during every image, which results in a random walk away from the mean behaviour. The actual atom number detected in the $ k $'th image is therefore
\begin{equation}\label{eq:TrueAtomNumber}
N_k = N_{k-1}p + \Delta N_k = N_0p^k + \sum_{j = 1}^{k} \Delta N_j.
\end{equation}
$ \Delta N_j $ are independent random variables with variance $ \Var(\Delta N_j) = N_j(1-p)$ and mean $ \mean{\Delta N_j}  = 0$ (see Eq.~\eqref{eq:stochLoss}).
This realizes a modified Poisson distribution displaced by the mean value to be centred around 0.
  Estimating the  atom number at the beginning of the imaging series from the $ k $'th image, $ \tilde{N}_{0,k} $, we obtain
\begin{equation}\label{eq:initAtomEst}
\tilde{N}_{0,k} = N_kp^{-k} = N_0 - p^{-k} \sum_{j = 1}^{k} \Delta N_j.
\end{equation}
The mean over all $ n $ images is thus
\begin{equation}\label{eq:initAtomAverageEst}
\tilde{N}_0 = \frac{1}{n} \sum_{k=1}^{n} \tilde{N}_{0,k}  = N_0 - \frac{1}{n} \sum_{k=1}^{n} p^{-k}\sum_{j = 1}^{k} \Delta N_j.
\end{equation}
The order of summation can be swapped since $ j \le n$ and $ k \ge j$. Defining $ \Delta \tilde{N}_0 = \tilde{N}_0 - N_0$, we obtain
\begin{equation}\label{eq:dN0}
\Delta \tilde{N}_0 = \frac{1}{n}\sum_{j=1}^{n} \Delta N_j \sum_{k=j}^{n} p^{-k} = \sum_{j=1}^{n} w_{n,j} \Delta N_j,
\end{equation} 
where the `weighting factor' is defined as \mbox{$ w_{n,j} =  \sum_{k=j}^{n} p^{-k}/n$.} The summation corresponds to a geometric series, and is thus given by
\begin{equation}\label{eq:weightFact}
w_{n,j} =  \sum_{k=j}^{n} \frac{p^{-k}}{n} = \frac{1}{n} \frac{p^{-j} - p^{-(n+1)}} {1-p^{-1}}.
\end{equation}
The variance of the mean due to the stochastic atom loss is then
\begin{equation}
\sigma_\mathrm{stoch}^2 =  \Var(\Delta \tilde{N}_0) = \sum_{j=1}^{n} w_{n,j}^2 \Var(\Delta N_j),
\end{equation}
and using the fact that $ \Var(\Delta N_j) = N_j(1-p)$,
\begin{equation}\label{eq:stochNoise}
\sigma_\mathrm{stoch}^2 = \sum_{j=1}^{n} w_{n,j}^2 N_j (1-p) = N_0(1-p) \sum_{j=1}^{n} w_{n,j}^2p^j.
\end{equation}
This shows that the stochastic atom loss noise for an average over $ n $ image is a factor $ \sum_{j=1}^{n} w_{n,j}^2p^j $ larger than the single image case. 
For a typical imaging series containing 50 images and an imaging pulse duration $t=0.66$~ms \cite{Gajdacz2016},  the stochastic atom loss noise is a factor of 108 larger than that of a  single image.

\bibliography{references2}

\begin{thebibliography}{31}%
\makeatletter
\providecommand \@ifxundefined [1]{%
 \@ifx{#1\undefined}
}%
\providecommand \@ifnum [1]{%
 \ifnum #1\expandafter \@firstoftwo
 \else \expandafter \@secondoftwo
 \fi
}%
\providecommand \@ifx [1]{%
 \ifx #1\expandafter \@firstoftwo
 \else \expandafter \@secondoftwo
 \fi
}%
\providecommand \natexlab [1]{#1}%
\providecommand \enquote  [1]{``#1''}%
\providecommand \bibnamefont  [1]{#1}%
\providecommand \bibfnamefont [1]{#1}%
\providecommand \citenamefont [1]{#1}%
\providecommand \href@noop [0]{\@secondoftwo}%
\providecommand \href [0]{\begingroup \@sanitize@url \@href}%
\providecommand \@href[1]{\@@startlink{#1}\@@href}%
\providecommand \@@href[1]{\endgroup#1\@@endlink}%
\providecommand \@sanitize@url [0]{\catcode `\\12\catcode `\$12\catcode
  `\&12\catcode `\#12\catcode `\^12\catcode `\_12\catcode `\%12\relax}%
\providecommand \@@startlink[1]{}%
\providecommand \@@endlink[0]{}%
\providecommand \url  [0]{\begingroup\@sanitize@url \@url }%
\providecommand \@url [1]{\endgroup\@href {#1}{\urlprefix }}%
\providecommand \urlprefix  [0]{URL }%
\providecommand \Eprint [0]{\href }%
\providecommand \doibase [0]{http://dx.doi.org/}%
\providecommand \selectlanguage [0]{\@gobble}%
\providecommand \bibinfo  [0]{\@secondoftwo}%
\providecommand \bibfield  [0]{\@secondoftwo}%
\providecommand \translation [1]{[#1]}%
\providecommand \BibitemOpen [0]{}%
\providecommand \bibitemStop [0]{}%
\providecommand \bibitemNoStop [0]{.\EOS\space}%
\providecommand \EOS [0]{\spacefactor3000\relax}%
\providecommand \BibitemShut  [1]{\csname bibitem#1\endcsname}%
\let\auto@bib@innerbib\@empty
\bibitem [{\citenamefont {Andrews}\ \emph {et~al.}(1996)\citenamefont
  {Andrews}, \citenamefont {Mewes}, \citenamefont {van Druten}, \citenamefont
  {Durfee}, \citenamefont {Kurn},\ and\ \citenamefont
  {Ketterle}}]{Andrews1996}%
  \BibitemOpen
  \bibfield  {author} {\bibinfo {author} {\bibfnamefont {M.~R.}\ \bibnamefont
  {Andrews}}, \bibinfo {author} {\bibfnamefont {M.-O.}\ \bibnamefont {Mewes}},
  \bibinfo {author} {\bibfnamefont {N.~J.}\ \bibnamefont {van Druten}},
  \bibinfo {author} {\bibfnamefont {D.~S.}\ \bibnamefont {Durfee}}, \bibinfo
  {author} {\bibfnamefont {D.~M.}\ \bibnamefont {Kurn}}, \ and\ \bibinfo
  {author} {\bibfnamefont {W.}~\bibnamefont {Ketterle}},\ }\href {\doibase
  10.1126/science.273.5271.84} {\bibfield  {journal} {\bibinfo  {journal}
  {Science (80-. ).}\ }\textbf {\bibinfo {volume} {273}},\ \bibinfo {pages}
  {84} (\bibinfo {year} {1996})}\BibitemShut {NoStop}%
\bibitem [{\citenamefont {Bradley}\ \emph
  {et~al.}(1997{\natexlab{a}})\citenamefont {Bradley}, \citenamefont
  {Sackett},\ and\ \citenamefont {Hulet}}]{Bradley1997}%
  \BibitemOpen
  \bibfield  {author} {\bibinfo {author} {\bibfnamefont {C.~C.}\ \bibnamefont
  {Bradley}}, \bibinfo {author} {\bibfnamefont {C.~A.}\ \bibnamefont
  {Sackett}}, \ and\ \bibinfo {author} {\bibfnamefont {R.~G.}\ \bibnamefont
  {Hulet}},\ }\href {\doibase 10.1103/PhysRevLett.78.985} {\bibfield  {journal}
  {\bibinfo  {journal} {Phys. Rev. Lett.}\ }\textbf {\bibinfo {volume} {78}},\
  \bibinfo {pages} {985} (\bibinfo {year} {1997}{\natexlab{a}})}\BibitemShut
  {NoStop}%
\bibitem [{\citenamefont {{Wigley}}\ \emph {et~al.}(2016)\citenamefont
  {{Wigley}}, \citenamefont {{Everitt}}, \citenamefont {{Hardman}},
  \citenamefont {{Sooriyabandara}}, \citenamefont {{Perumbil}}, \citenamefont
  {{Close}}, \citenamefont {{Robins}},\ and\ \citenamefont
  {{Kuhn}}}]{2016arXiv160104425W}%
  \BibitemOpen
  \bibfield  {author} {\bibinfo {author} {\bibfnamefont {P.}~\bibnamefont
  {{Wigley}}}, \bibinfo {author} {\bibfnamefont {P.}~\bibnamefont {{Everitt}}},
  \bibinfo {author} {\bibfnamefont {K.}~\bibnamefont {{Hardman}}}, \bibinfo
  {author} {\bibfnamefont {M.}~\bibnamefont {{Sooriyabandara}}}, \bibinfo
  {author} {\bibfnamefont {M.}~\bibnamefont {{Perumbil}}}, \bibinfo {author}
  {\bibfnamefont {J.}~\bibnamefont {{Close}}}, \bibinfo {author} {\bibfnamefont
  {N.}~\bibnamefont {{Robins}}}, \ and\ \bibinfo {author} {\bibfnamefont
  {C.}~\bibnamefont {{Kuhn}}},\ }\href@noop {} {\bibfield  {journal} {\bibinfo
  {journal} {ArXiv e-prints}\ } (\bibinfo {year} {2016})},\ \Eprint
  {http://arxiv.org/abs/1601.04425} {arXiv:1601.04425 [physics.atom-ph]}
  \BibitemShut {NoStop}%
\bibitem [{\citenamefont {Andrews}\ \emph {et~al.}(1997)\citenamefont
  {Andrews}, \citenamefont {Kurn}, \citenamefont {Miesner}, \citenamefont
  {Durfee}, \citenamefont {Townsend}, \citenamefont {Inouye},\ and\
  \citenamefont {Ketterle}}]{PhysRevLett.79.553}%
  \BibitemOpen
  \bibfield  {author} {\bibinfo {author} {\bibfnamefont {M.~R.}\ \bibnamefont
  {Andrews}}, \bibinfo {author} {\bibfnamefont {D.~M.}\ \bibnamefont {Kurn}},
  \bibinfo {author} {\bibfnamefont {H.-J.}\ \bibnamefont {Miesner}}, \bibinfo
  {author} {\bibfnamefont {D.~S.}\ \bibnamefont {Durfee}}, \bibinfo {author}
  {\bibfnamefont {C.~G.}\ \bibnamefont {Townsend}}, \bibinfo {author}
  {\bibfnamefont {S.}~\bibnamefont {Inouye}}, \ and\ \bibinfo {author}
  {\bibfnamefont {W.}~\bibnamefont {Ketterle}},\ }\href {\doibase
  10.1103/PhysRevLett.79.553} {\bibfield  {journal} {\bibinfo  {journal} {Phys.
  Rev. Lett.}\ }\textbf {\bibinfo {volume} {79}},\ \bibinfo {pages} {553}
  (\bibinfo {year} {1997})}\BibitemShut {NoStop}%
\bibitem [{\citenamefont {Matthews}\ \emph {et~al.}(1999)\citenamefont
  {Matthews}, \citenamefont {Anderson}, \citenamefont {Haljan}, \citenamefont
  {Hall}, \citenamefont {Wieman},\ and\ \citenamefont
  {Cornell}}]{PhysRevLett.83.2498}%
  \BibitemOpen
  \bibfield  {author} {\bibinfo {author} {\bibfnamefont {M.~R.}\ \bibnamefont
  {Matthews}}, \bibinfo {author} {\bibfnamefont {B.~P.}\ \bibnamefont
  {Anderson}}, \bibinfo {author} {\bibfnamefont {P.~C.}\ \bibnamefont
  {Haljan}}, \bibinfo {author} {\bibfnamefont {D.~S.}\ \bibnamefont {Hall}},
  \bibinfo {author} {\bibfnamefont {C.~E.}\ \bibnamefont {Wieman}}, \ and\
  \bibinfo {author} {\bibfnamefont {E.~A.}\ \bibnamefont {Cornell}},\ }\href
  {\doibase 10.1103/PhysRevLett.83.2498} {\bibfield  {journal} {\bibinfo
  {journal} {Phys. Rev. Lett.}\ }\textbf {\bibinfo {volume} {83}},\ \bibinfo
  {pages} {2498} (\bibinfo {year} {1999})}\BibitemShut {NoStop}%
\bibitem [{\citenamefont {Bons}\ \emph {et~al.}(2016)\citenamefont {Bons},
  \citenamefont {de~Haas}, \citenamefont {de~Jong}, \citenamefont {Groot},\
  and\ \citenamefont {van~der Straten}}]{PhysRevLett.116.173602}%
  \BibitemOpen
  \bibfield  {author} {\bibinfo {author} {\bibfnamefont {P.~C.}\ \bibnamefont
  {Bons}}, \bibinfo {author} {\bibfnamefont {R.}~\bibnamefont {de~Haas}},
  \bibinfo {author} {\bibfnamefont {D.}~\bibnamefont {de~Jong}}, \bibinfo
  {author} {\bibfnamefont {A.}~\bibnamefont {Groot}}, \ and\ \bibinfo {author}
  {\bibfnamefont {P.}~\bibnamefont {van~der Straten}},\ }\href {\doibase
  10.1103/PhysRevLett.116.173602} {\bibfield  {journal} {\bibinfo  {journal}
  {Phys. Rev. Lett.}\ }\textbf {\bibinfo {volume} {116}},\ \bibinfo {pages}
  {173602} (\bibinfo {year} {2016})}\BibitemShut {NoStop}%
\bibitem [{\citenamefont {Gajdacz}\ \emph {et~al.}(2013)\citenamefont
  {Gajdacz}, \citenamefont {Pedersen}, \citenamefont {M{\o}rch}, \citenamefont
  {Hilliard}, \citenamefont {Arlt},\ and\ \citenamefont
  {Sherson}}]{Gajdacz2013}%
  \BibitemOpen
  \bibfield  {author} {\bibinfo {author} {\bibfnamefont {M.}~\bibnamefont
  {Gajdacz}}, \bibinfo {author} {\bibfnamefont {P.~L.}\ \bibnamefont
  {Pedersen}}, \bibinfo {author} {\bibfnamefont {T.}~\bibnamefont {M{\o}rch}},
  \bibinfo {author} {\bibfnamefont {A.~J.}\ \bibnamefont {Hilliard}}, \bibinfo
  {author} {\bibfnamefont {J.}~\bibnamefont {Arlt}}, \ and\ \bibinfo {author}
  {\bibfnamefont {J.~F.}\ \bibnamefont {Sherson}},\ }\href@noop {} {\bibfield
  {journal} {\bibinfo  {journal} {Rev. Sci. Instrum.}\ }\textbf {\bibinfo
  {volume} {84}} (\bibinfo {year} {2013})},\ \Eprint
  {http://arxiv.org/abs/1301.3018} {arXiv:1301.3018} \BibitemShut {NoStop}%
\bibitem [{\citenamefont {Bradley}\ \emph
  {et~al.}(1997{\natexlab{b}})\citenamefont {Bradley}, \citenamefont
  {Sackett},\ and\ \citenamefont {Hulet}}]{PhysRevA.55.3951}%
  \BibitemOpen
  \bibfield  {author} {\bibinfo {author} {\bibfnamefont {C.~C.}\ \bibnamefont
  {Bradley}}, \bibinfo {author} {\bibfnamefont {C.~A.}\ \bibnamefont
  {Sackett}}, \ and\ \bibinfo {author} {\bibfnamefont {R.~G.}\ \bibnamefont
  {Hulet}},\ }\href {\doibase 10.1103/PhysRevA.55.3951} {\bibfield  {journal}
  {\bibinfo  {journal} {Phys. Rev. A}\ }\textbf {\bibinfo {volume} {55}},\
  \bibinfo {pages} {3951} (\bibinfo {year} {1997}{\natexlab{b}})}\BibitemShut
  {NoStop}%
\bibitem [{\citenamefont {Kaminski}\ \emph {et~al.}(2012)\citenamefont
  {Kaminski}, \citenamefont {Kampel}, \citenamefont {Steenstrup}, \citenamefont
  {Griesmaier}, \citenamefont {Polzik},\ and\ \citenamefont
  {M\"uller}}]{Kaminski2012}%
  \BibitemOpen
  \bibfield  {author} {\bibinfo {author} {\bibfnamefont {F.}~\bibnamefont
  {Kaminski}}, \bibinfo {author} {\bibfnamefont {N.}~\bibnamefont {Kampel}},
  \bibinfo {author} {\bibfnamefont {M.}~\bibnamefont {Steenstrup}}, \bibinfo
  {author} {\bibfnamefont {A.}~\bibnamefont {Griesmaier}}, \bibinfo {author}
  {\bibfnamefont {E.}~\bibnamefont {Polzik}}, \ and\ \bibinfo {author}
  {\bibfnamefont {J.}~\bibnamefont {M\"uller}},\ }\href {\doibase
  10.1140/epjd/e2012-30038-0} {\bibfield  {journal} {\bibinfo  {journal} {Eur.
  Phys. J. D}\ }\textbf {\bibinfo {volume} {66}},\ \bibinfo {pages} {1}
  (\bibinfo {year} {2012})}\BibitemShut {NoStop}%
\bibitem [{\citenamefont {Hope}\ and\ \citenamefont {Close}(2004)}]{Hope2004}%
  \BibitemOpen
  \bibfield  {author} {\bibinfo {author} {\bibfnamefont {J.~J.}\ \bibnamefont
  {Hope}}\ and\ \bibinfo {author} {\bibfnamefont {J.~D.}\ \bibnamefont
  {Close}},\ }\href {\doibase 10.1103/PhysRevLett.93.180402} {\bibfield
  {journal} {\bibinfo  {journal} {Phys. Rev. Lett.}\ }\textbf {\bibinfo
  {volume} {93}},\ \bibinfo {pages} {180402} (\bibinfo {year}
  {2004})}\BibitemShut {NoStop}%
\bibitem [{Note1()}]{Note1}%
  \BibitemOpen
  \bibinfo {note} {The Faraday coefficient also depends on the total angular
  momentum $F$. See \cite {Gajdacz2013} for details on the calculation of
  $c_\protect \textrm {F}$ and how it changes for the experimentally relevant
  values of alkali atom nuclear spin $I=3/2$ and $7/2$.}\BibitemShut {Stop}%
\bibitem [{\citenamefont {Ketterle}\ \emph {et~al.}(1999)\citenamefont
  {Ketterle}, \citenamefont {Durfee},\ and\ \citenamefont
  {Stamper-Kurn}}]{Ketterle}%
  \BibitemOpen
  \bibfield  {author} {\bibinfo {author} {\bibfnamefont {W.}~\bibnamefont
  {Ketterle}}, \bibinfo {author} {\bibfnamefont {D.}~\bibnamefont {Durfee}}, \
  and\ \bibinfo {author} {\bibfnamefont {D.}~\bibnamefont {Stamper-Kurn}},\
  }in\ \href@noop {} {\emph {\bibinfo {booktitle} {{BOSE-EINSTEIN CONDENSATION
  IN ATOMIC GASES}}}},\ \bibinfo {series} {{PROCEEDINGS OF THE INTERNATIONAL
  SCHOOL OF PHYSICS ENRICO FERMI}}, Vol.\ \bibinfo {volume} {{140}},\ \bibinfo
  {editor} {edited by\ \bibinfo {editor} {\bibnamefont {{Inguscio, M and
  Stringari, S and Wieman, CE}}}}\ (\bibinfo {organization} {{Inst Sch Phys
  Enrico Fermi}},\ \bibinfo {year} {{1999}})\ pp.\ \bibinfo {pages}
  {{67--176}},\ \bibinfo {note} {{International School of Physics Enrico Fermi
  on Bose-Einstein Condensation in Atomic Gases, VARENNA LAKE COMO, ITALY, JUL
  07-17, 1998}}\BibitemShut {NoStop}%
\bibitem [{\citenamefont {Lewandowski}\ \emph {et~al.}(2003)\citenamefont
  {Lewandowski}, \citenamefont {Harber}, \citenamefont {Whitaker},\ and\
  \citenamefont {Cornell}}]{Lewandowski2003}%
  \BibitemOpen
  \bibfield  {author} {\bibinfo {author} {\bibfnamefont {H.~J.}\ \bibnamefont
  {Lewandowski}}, \bibinfo {author} {\bibfnamefont {D.~M.}\ \bibnamefont
  {Harber}}, \bibinfo {author} {\bibfnamefont {D.~L.}\ \bibnamefont
  {Whitaker}}, \ and\ \bibinfo {author} {\bibfnamefont {E.~A.}\ \bibnamefont
  {Cornell}},\ }\href {\doibase 10.1023/A:1024800600621} {\bibfield  {journal}
  {\bibinfo  {journal} {Journal of Low Temperature Physics}\ }\textbf {\bibinfo
  {volume} {132}},\ \bibinfo {pages} {309} (\bibinfo {year}
  {2003})}\BibitemShut {NoStop}%
\bibitem [{\citenamefont {Bertelsen}\ \emph {et~al.}(2007)\citenamefont
  {Bertelsen}, \citenamefont {Andersen}, \citenamefont {Mai},\ and\
  \citenamefont {Budde}}]{Bertelsen2007}%
  \BibitemOpen
  \bibfield  {author} {\bibinfo {author} {\bibfnamefont {J.~F.}\ \bibnamefont
  {Bertelsen}}, \bibinfo {author} {\bibfnamefont {H.~K.}\ \bibnamefont
  {Andersen}}, \bibinfo {author} {\bibfnamefont {S.}~\bibnamefont {Mai}}, \
  and\ \bibinfo {author} {\bibfnamefont {M.}~\bibnamefont {Budde}},\ }\href
  {\doibase 10.1103/PhysRevA.75.013404} {\bibfield  {journal} {\bibinfo
  {journal} {Phys. Rev. A}\ }\textbf {\bibinfo {volume} {75}},\ \bibinfo
  {pages} {013404} (\bibinfo {year} {2007})}\BibitemShut {NoStop}%
\bibitem [{\citenamefont {Esslinger}\ \emph {et~al.}(1998)\citenamefont
  {Esslinger}, \citenamefont {Bloch},\ and\ \citenamefont
  {H\"ansch}}]{Esslinger1998}%
  \BibitemOpen
  \bibfield  {author} {\bibinfo {author} {\bibfnamefont {T.}~\bibnamefont
  {Esslinger}}, \bibinfo {author} {\bibfnamefont {I.}~\bibnamefont {Bloch}}, \
  and\ \bibinfo {author} {\bibfnamefont {T.~W.}\ \bibnamefont {H\"ansch}},\
  }\href {\doibase 10.1103/PhysRevA.58.R2664} {\bibfield  {journal} {\bibinfo
  {journal} {Phys. Rev. A}\ }\textbf {\bibinfo {volume} {58}},\ \bibinfo
  {pages} {R2664} (\bibinfo {year} {1998})}\BibitemShut {NoStop}%
\bibitem [{\citenamefont {Schünemann}\ \emph {et~al.}(1999)\citenamefont
  {Schünemann}, \citenamefont {Engler}, \citenamefont {Grimm}, \citenamefont
  {Weidemüller},\ and\ \citenamefont {Zielonkowski}}]{Schuenemann1999}%
  \BibitemOpen
  \bibfield  {author} {\bibinfo {author} {\bibfnamefont {U.}~\bibnamefont
  {Schünemann}}, \bibinfo {author} {\bibfnamefont {H.}~\bibnamefont {Engler}},
  \bibinfo {author} {\bibfnamefont {R.}~\bibnamefont {Grimm}}, \bibinfo
  {author} {\bibfnamefont {M.}~\bibnamefont {Weidemüller}}, \ and\ \bibinfo
  {author} {\bibfnamefont {M.}~\bibnamefont {Zielonkowski}},\ }\href {\doibase
  http://dx.doi.org/10.1063/1.1149573} {\bibfield  {journal} {\bibinfo
  {journal} {Review of Scientific Instruments}\ }\textbf {\bibinfo {volume}
  {70}},\ \bibinfo {pages} {242} (\bibinfo {year} {1999})}\BibitemShut
  {NoStop}%
\bibitem [{Note2()}]{Note2}%
  \BibitemOpen
  \bibinfo {note} {Due to the size of the beam relative to the cloud and the
  modest peak rotation angles, the reduction in photodiode signal due to the
  atoms is negligible.}\BibitemShut {Stop}%
\bibitem [{\citenamefont {Reinaudi}\ \emph {et~al.}(2007)\citenamefont
  {Reinaudi}, \citenamefont {Lahaye}, \citenamefont {Wang},\ and\ \citenamefont
  {Gu\'{e}ry-Odelin}}]{Reinaudi2007}%
  \BibitemOpen
  \bibfield  {author} {\bibinfo {author} {\bibfnamefont {G.}~\bibnamefont
  {Reinaudi}}, \bibinfo {author} {\bibfnamefont {T.}~\bibnamefont {Lahaye}},
  \bibinfo {author} {\bibfnamefont {Z.}~\bibnamefont {Wang}}, \ and\ \bibinfo
  {author} {\bibfnamefont {D.}~\bibnamefont {Gu\'{e}ry-Odelin}},\ }\href
  {\doibase 10.1364/OL.32.003143} {\bibfield  {journal} {\bibinfo  {journal}
  {Opt. Lett.}\ }\textbf {\bibinfo {volume} {32}},\ \bibinfo {pages} {3143}
  (\bibinfo {year} {2007})}\BibitemShut {NoStop}%
\bibitem [{\citenamefont {Muessel}\ \emph {et~al.}(2013)\citenamefont
  {Muessel}, \citenamefont {Strobel}, \citenamefont {Joos}, \citenamefont
  {Nicklas}, \citenamefont {Stroescu}, \citenamefont {Tomkovi{\v{c}}},
  \citenamefont {Hume},\ and\ \citenamefont {Oberthaler}}]{Muessel2013}%
  \BibitemOpen
  \bibfield  {author} {\bibinfo {author} {\bibfnamefont {W.}~\bibnamefont
  {Muessel}}, \bibinfo {author} {\bibfnamefont {H.}~\bibnamefont {Strobel}},
  \bibinfo {author} {\bibfnamefont {M.}~\bibnamefont {Joos}}, \bibinfo {author}
  {\bibfnamefont {E.}~\bibnamefont {Nicklas}}, \bibinfo {author} {\bibfnamefont
  {I.}~\bibnamefont {Stroescu}}, \bibinfo {author} {\bibfnamefont
  {J.}~\bibnamefont {Tomkovi{\v{c}}}}, \bibinfo {author} {\bibfnamefont
  {D.~B.}\ \bibnamefont {Hume}}, \ and\ \bibinfo {author} {\bibfnamefont
  {M.~K.}\ \bibnamefont {Oberthaler}},\ }\href {\doibase
  10.1007/s00340-013-5553-8} {\bibfield  {journal} {\bibinfo  {journal}
  {Applied Physics B}\ }\textbf {\bibinfo {volume} {113}},\ \bibinfo {pages}
  {69} (\bibinfo {year} {2013})}\BibitemShut {NoStop}%
\bibitem [{\citenamefont {van Kampen}(2007)}]{Kampen2007}%
  \BibitemOpen
  \bibfield  {author} {\bibinfo {author} {\bibfnamefont {N.~G.}\ \bibnamefont
  {van Kampen}},\ }\href@noop {} {\emph {\bibinfo {title} {Stochastic Processes
  in Physics and Chemistry}}},\ \bibinfo {edition} {3rd}\ ed.\ (\bibinfo
  {publisher} {Elsevier, Amsterdam},\ \bibinfo {year} {2007})\BibitemShut
  {NoStop}%
\bibitem [{\citenamefont {Gajdacz}\ \emph {et~al.}(2016)\citenamefont
  {Gajdacz}, \citenamefont {Hilliard}, \citenamefont {Kristensen},
  \citenamefont {Pedersen}, \citenamefont {Klempt}, \citenamefont {Arlt},\ and\
  \citenamefont {Sherson}}]{Gajdacz2016}%
  \BibitemOpen
  \bibfield  {author} {\bibinfo {author} {\bibfnamefont {M.}~\bibnamefont
  {Gajdacz}}, \bibinfo {author} {\bibfnamefont {A.~J.}\ \bibnamefont
  {Hilliard}}, \bibinfo {author} {\bibfnamefont {M.~A.}\ \bibnamefont
  {Kristensen}}, \bibinfo {author} {\bibfnamefont {P.~L.}\ \bibnamefont
  {Pedersen}}, \bibinfo {author} {\bibfnamefont {C.}~\bibnamefont {Klempt}},
  \bibinfo {author} {\bibfnamefont {J.~J.}\ \bibnamefont {Arlt}}, \ and\
  \bibinfo {author} {\bibfnamefont {J.~F.}\ \bibnamefont {Sherson}},\ }\href
  {\doibase 10.1103/PhysRevLett.117.073604} {\bibfield  {journal} {\bibinfo
  {journal} {Phys. Rev. Lett.}\ }\textbf {\bibinfo {volume} {117}},\ \bibinfo
  {pages} {073604} (\bibinfo {year} {2016})}\BibitemShut {NoStop}%
\bibitem [{\citenamefont {Hilliard}\ \emph {et~al.}(2015)\citenamefont
  {Hilliard}, \citenamefont {Fung}, \citenamefont {Sompet}, \citenamefont
  {Carpentier},\ and\ \citenamefont {Andersen}}]{Hilliard2015}%
  \BibitemOpen
  \bibfield  {author} {\bibinfo {author} {\bibfnamefont {A.~J.}\ \bibnamefont
  {Hilliard}}, \bibinfo {author} {\bibfnamefont {Y.~H.}\ \bibnamefont {Fung}},
  \bibinfo {author} {\bibfnamefont {P.}~\bibnamefont {Sompet}}, \bibinfo
  {author} {\bibfnamefont {A.~V.}\ \bibnamefont {Carpentier}}, \ and\ \bibinfo
  {author} {\bibfnamefont {M.~F.}\ \bibnamefont {Andersen}},\ }\href {\doibase
  10.1103/PhysRevA.91.053414} {\bibfield  {journal} {\bibinfo  {journal} {Phys.
  Rev. A}\ }\textbf {\bibinfo {volume} {91}},\ \bibinfo {pages} {053414}
  (\bibinfo {year} {2015})}\BibitemShut {NoStop}%
\bibitem [{\citenamefont {Pethick}\ and\ \citenamefont
  {Smith}(2008)}]{Pethick2008}%
  \BibitemOpen
  \bibfield  {author} {\bibinfo {author} {\bibfnamefont {C.~J.}\ \bibnamefont
  {Pethick}}\ and\ \bibinfo {author} {\bibfnamefont {H.}~\bibnamefont
  {Smith}},\ }\href@noop {} {\emph {\bibinfo {title} {{Bose-Einstein}
  Condensation in Dilute Gases}}},\ \bibinfo {edition} {second edition}\ ed.\
  (\bibinfo  {publisher} {Cambridge},\ \bibinfo {year} {2008})\BibitemShut
  {NoStop}%
\bibitem [{Note3()}]{Note3}%
  \BibitemOpen
  \bibinfo {note} {The Faraday rotation angles can also be reconstructed from
  the evaluated signal through \unhbox \voidb@x \hbox {$ \theta = \protect
  \qopname \relax o{arcsin}(\protect \sqrt {S})$} \cite {2016arXiv160702934B},
  but this approach can be problematic. First, since the baseline is
  subtracted, the pixel values where the signal is low can become negative,
  which complicates the evaluation of the square root. Secondly, the steep
  derivative of the square root for small arguments amplifies the noise in
  pixels with low signal.}\BibitemShut {Stop}%
\bibitem [{\citenamefont {Ketterle}\ and\ \citenamefont
  {Druten}(1996)}]{Ketterle1996}%
  \BibitemOpen
  \bibfield  {author} {\bibinfo {author} {\bibfnamefont {W.}~\bibnamefont
  {Ketterle}}\ and\ \bibinfo {author} {\bibfnamefont {N.~V.}\ \bibnamefont
  {Druten}}\ }(\bibinfo  {publisher} {Academic Press},\ \bibinfo {year}
  {1996})\ pp.\ \bibinfo {pages} {181 -- 236}\BibitemShut {NoStop}%
\bibitem [{\citenamefont {Robbins}\ and\ \citenamefont
  {Hadwen}(2003)}]{Robbins2003}%
  \BibitemOpen
  \bibfield  {author} {\bibinfo {author} {\bibfnamefont {M.~S.}\ \bibnamefont
  {Robbins}}\ and\ \bibinfo {author} {\bibfnamefont {B.~J.}\ \bibnamefont
  {Hadwen}},\ }\href {\doibase 10.1109/TED.2003.813462} {\bibfield  {journal}
  {\bibinfo  {journal} {IEEE Trans. Electron Devices}\ }\textbf {\bibinfo
  {volume} {50}},\ \bibinfo {pages} {1227} (\bibinfo {year}
  {2003})}\BibitemShut {NoStop}%
\bibitem [{\citenamefont {Hynecek}\ \emph {et~al.}(2003)\citenamefont
  {Hynecek}, \citenamefont {Member},\ and\ \citenamefont
  {Nishiwaki}}]{Hynecek2003}%
  \BibitemOpen
  \bibfield  {author} {\bibinfo {author} {\bibfnamefont {J.}~\bibnamefont
  {Hynecek}}, \bibinfo {author} {\bibfnamefont {S.}~\bibnamefont {Member}}, \
  and\ \bibinfo {author} {\bibfnamefont {T.}~\bibnamefont {Nishiwaki}},\
  }\href@noop {} {\bibfield  {journal} {\bibinfo  {journal} {IEEE Trans.
  Electron Devices}\ }\textbf {\bibinfo {volume} {50}},\ \bibinfo {pages} {239}
  (\bibinfo {year} {2003})}\BibitemShut {NoStop}%
\bibitem [{\citenamefont {Hume}\ \emph {et~al.}(2013)\citenamefont {Hume},
  \citenamefont {Stroescu}, \citenamefont {Joos}, \citenamefont {Muessel},
  \citenamefont {Strobel},\ and\ \citenamefont {Oberthaler}}]{Hume2013}%
  \BibitemOpen
  \bibfield  {author} {\bibinfo {author} {\bibfnamefont {D.~B.}\ \bibnamefont
  {Hume}}, \bibinfo {author} {\bibfnamefont {I.}~\bibnamefont {Stroescu}},
  \bibinfo {author} {\bibfnamefont {M.}~\bibnamefont {Joos}}, \bibinfo {author}
  {\bibfnamefont {W.}~\bibnamefont {Muessel}}, \bibinfo {author} {\bibfnamefont
  {H.}~\bibnamefont {Strobel}}, \ and\ \bibinfo {author} {\bibfnamefont
  {M.~K.}\ \bibnamefont {Oberthaler}},\ }\href {\doibase
  10.1103/PhysRevLett.111.253001} {\bibfield  {journal} {\bibinfo  {journal}
  {Phys. Rev. Lett.}\ }\textbf {\bibinfo {volume} {111}},\ \bibinfo {pages}
  {253001} (\bibinfo {year} {2013})}\BibitemShut {NoStop}%
\bibitem [{\citenamefont {Efron}(1979)}]{Efron1979}%
  \BibitemOpen
  \bibfield  {author} {\bibinfo {author} {\bibfnamefont {B.}~\bibnamefont
  {Efron}},\ }\href {\doibase 10.1214/aos/1176344552} {\bibfield  {journal}
  {\bibinfo  {journal} {Ann. Statist.}\ }\textbf {\bibinfo {volume} {7}},\
  \bibinfo {pages} {1} (\bibinfo {year} {1979})}\BibitemShut {NoStop}%
\bibitem [{\citenamefont {Steck}(2015)}]{SteckRB87}%
  \BibitemOpen
  \bibfield  {author} {\bibinfo {author} {\bibfnamefont {D.}~\bibnamefont
  {Steck}},\ }\href@noop {} {\enquote {\bibinfo {title} {{Rubidium 87 D Line
  Data}},}\ }\bibinfo {howpublished} {Online} (\bibinfo {year}
  {2015})\BibitemShut {NoStop}%
\bibitem [{\citenamefont {{Bason}}\ \emph {et~al.}(2016)\citenamefont
  {{Bason}}, \citenamefont {{Heck}}, \citenamefont {{Napolitano}},
  \citenamefont {{El{\'{\i}}asson}}, \citenamefont {{M{\"u}ller}},
  \citenamefont {{Thorsen}}, \citenamefont {{Zhang}}, \citenamefont {{Arlt}},\
  and\ \citenamefont {{Sherson}}}]{2016arXiv160702934B}%
  \BibitemOpen
  \bibfield  {author} {\bibinfo {author} {\bibfnamefont {M.~G.}\ \bibnamefont
  {{Bason}}}, \bibinfo {author} {\bibfnamefont {R.}~\bibnamefont {{Heck}}},
  \bibinfo {author} {\bibfnamefont {M.}~\bibnamefont {{Napolitano}}}, \bibinfo
  {author} {\bibfnamefont {O.}~\bibnamefont {{El{\'{\i}}asson}}}, \bibinfo
  {author} {\bibfnamefont {R.}~\bibnamefont {{M{\"u}ller}}}, \bibinfo {author}
  {\bibfnamefont {A.}~\bibnamefont {{Thorsen}}}, \bibinfo {author}
  {\bibfnamefont {W.-Z.}\ \bibnamefont {{Zhang}}}, \bibinfo {author}
  {\bibfnamefont {J.}~\bibnamefont {{Arlt}}}, \ and\ \bibinfo {author}
  {\bibfnamefont {J.~F.}\ \bibnamefont {{Sherson}}},\ }\href@noop {} {\bibfield
   {journal} {\bibinfo  {journal} {ArXiv e-prints}\ } (\bibinfo {year}
  {2016})},\ \Eprint {http://arxiv.org/abs/1607.02934} {arXiv:1607.02934
  [quant-ph]} \BibitemShut {NoStop}%
\end{thebibliography}%

\end{document}